# Chiral magnetization textures stabilized by the Dzyaloshinskii-Moriya interaction during spin-orbit torque switching

N. Perez[1,a)], E. Martinez[1], L. Torres[1], S.-H. Woo[2], S. Emori[2], and G. S. D. Beach[2]

[1]*Departamento de Física Aplicada, Universidad de Salamanca, Plaza de los Caidos s/n, Salamanca 37008, Spain.*

[2]*Department of Materials Science and Engineering, Massachusetts Institute of Technology, Cambridge, Massachusetts 02139, USA.*

We study the effect of the Dzyaloshinskii-Moriya interaction (DMI) on current-induced magnetic switching of a perpendicularly magnetized heavy-metal/ferromagnet/oxide trilayer both experimentally and through micromagnetic simulations. We report the generation of stable helical magnetization stripes for a sufficiently large DMI strength in the switching region, giving rise to intermediate states in the magnetization confirming the essential role of the DMI on switching processes. We compare the simulation and experimental results to a macrospin model, showing the need for a micromagnetic approach. The influence of the temperature on the switching is also discussed.

---

[a)] Author to whom correspondence should be addressed. Electronic mail: noelpg@usal.es

It has been recently reported that the magnetization of perpendicular anisotropy heavy-metal/ferromagnet/oxide trilayers can be switched by in-plane injected current[1-5]. This current induced switching has been attributed to so called spin-orbit torques (SOTs) due to the spin-Hall effect (SHE)[2-6] and the Rashba effect[1,7-9], arising from spin-orbit coupling and broken inversion symmetry at the heavy-metal/ferromagnet interface. SOT switching has been studied theoretically[6] and experimentally for Pt/Co/AlO$_x$[1,3,7,10], Pt/Co/Pt[11], Ta/CoFeB/MgO[12], Pt/CoFe/MgO[5] and Ta/CoFe/MgO[5] nanodots and nanostrips[4,5,13,14], and may provide an efficient means to operate ferromagnetic memories[15] such as SOT-MRAM or racetrack memories[16]. More recently, it has been shown that the Dzyaloshinskii-Moriya interaction (DMI) plays an important role in domain wall (DW) motion in such materials[5,17-20] and may also be important in SOT switching[21].

The SHE in the heavy metal layer deflects electrons with opposite spin in opposite directions, giving rise to a spin current that is injected into the ferromagnet, exerting a Slonczewski-like torque on the ferromagnet magnetization[3]. The Rashba effect leads to a current-induced effective field parallel to the interface and transverse to the current, and may additionally generate a Slonczewski-like torque. On the other hand, the DMI[17-19] acts as an antisymmetric exchange interaction that promotes non-uniform magnetization in materials where inversion symmetry is broken due to lack of a lattice inversion center or to the presence of surfaces or interfaces. The DMI directly competes with the exchange interaction (which tends to maintain the magnetization uniform), and when strong enough encourages the formation of rotational magnetic textures of definite chirality such as chiral DWs[1,20], skyrmions[22], or spin helixes[23]. SOT switching has usually been interpreted in the framework of a simple macrospin model, but recent micromagnetic simulations show that the reversal process is in fact highly non-uniform[24]. In this case, it can be anticipated that strong DMI should significantly impact the inhomogeneous reversal process, necessitating a full micromagnetic model to provide meaningful interpretation of experiments.

Here we study SOT switching experimentally and through a micromagnetic model that includes the Slonczewski-like and field-like SOTs, DMI and finite temperature. Although the Slonczewski-like SOT drives magnetization switching, the field-like SOT and DMI play key roles in the switching process and switching efficiency. We show that strong DMI leads to metastable chiral intermediate states and introduces a stochastic nature to the switching process. A strong field-like torque, by contrast, promotes quasi-uniform, deterministic switching and significantly lowers the switching threshold.

Experiments focus on perpendicular anisotropy Ta(3nm)/Pt(3nm)/CoFe(0.6nm)/MgO(1.8nm)/Ta(2nm) and Ta(5nm)/CoFe(0.6nm)/MgO(1.8nm)/Ta(2nm) films sputter-deposited onto $Si/SiO_2$ substrates and patterned into 1200nm-wide Hall crosses (Fig. 1a). See Ref. 5 for sample details. Strong Slonczewski-like torques have been observed in Pt/CoFe and Ta/CoFe stacks[5] and attributed to large spin Hall angles[2,3] in Pt and Ta. At the same time, experiments indicate[5,25] Pt/CoFe exhibits strong DMI and a relatively weak field-like SOT, while Ta/CoFe exhibits weak DMI and a strong field-like SOT. These materials thus provide an opportunity to independently examine the roles of these effects on SOT switching.

The Slonczewski-like torque in heavy-metal/ferromagnet bilayers is described by an effective field $\mathbf{B}_{SL} \propto \mathbf{m} \times (\hat{\mathbf{z}} \times \mathbf{J})$, where $\mathbf{m} = \mathbf{M}/M_S$ is the normalized magnetization, $\hat{\mathbf{z}}$ is film normal, and $\mathbf{J}$ is the current density. A small in-plane field $B_x$ along the current axis (Fig. 1b) enables current-induced switching of the z-component of magnetization between $m_z \approx \pm 1$. Figures 1c,d show current-induced switching in Pt/CoFe and Ta/CoFe devices with $B_x = 50$ mT for Pt/CoFe and 10 mT for Ta/CoFe. For each point, a 250ms current pulse was injected, then $m_z$ was measured from the anomalous Hall effect (AHE) voltage using a 400Hz low-amplitude (~$10^9$ A/m$^2$) AC sense current and a lock-in amplifier. $m_z$ was measured successively in this fashion by stepping through a range of current pulse amplitudes. The switching polarities are opposite for Pt/CoFe and Ta/CoFe and are consistent with positive and negative spin Hall angles, respectively. The switching polarities reversed as expected when the direction of $B_x$ was reversed.

Switching in Pt/CoFe exhibited a stochastic behavior, leading to intermediate states near the switching threshold and occasional switch-backs at higher current densities (Fig. 1c). Such behavior was commonly observed in multiple nominally-identical devices, each measured several times. By contrast, sharp switching was consistently observed in all Ta/CoFe devices, with an absence of intermediate states (Fig. 1d).

To exclude the possibility that DW pinning or geometrical features in the Hall cross region contribute to the stochastic behavior, we used the polar magneto-optical Kerr effect (pMOKE) to probe $m_z$ along a straight region of the strip (Fig. 1a). Figures 2a,b present switching phase diagrams for Ta/CoFe and Pt/CoFe, respectively. For each set ($B_x, J$) the sample was initialized in the down state with a perpendicular field pulse. Then a 50ns current pulse was injected, after which $m_z$ was probed with pMOKE. The color scale shows $m_z$ obtained from the average of ten switching cycles. For Pt/CoFe, complete switching was observed only for $B_x > 8$ mT, at any $J$, above which the switching current threshold decreased approximitely linearly with $B_x$. A finite transition region between non-switching and switching is observed along the entire switching boundary. For Ta/CoFe, switching occurs at significantly lower $J$ and $B_x$ and the switching boundary is sharp (Fig. 2b).

Figures 2c,d show for Pt/CoFe the mean $m_z$ after current injection, averaged over 50 switching cycles and plotted versus $J$ (Fig. 2c) and $B_x$ (Fig. 2d) with higher step resolution than in Fig. 2a. From these figures, the finite breadth of the intermediate switching region is clear. Figure 2e shows a histogram of $m_z$ after current-pulse injection, for 50 individual switching cycles, with $J$, $B_x$ on the switching boundary. The data indicate 100% probability to switch into a state intermediate between 'up' and 'down', suggesting a stable multidomain configuration. Figures 2c,d show that destabilization of the uniform state occurs sharply at a lower threshold $B_x$ and $J$, and a second threshold at larger $B_x$ and $J$ exists whereupon nearly complete reversal occurs. For Ta/CoFe, by contrast, only a single threshold is observed (Fig. 2f) and intermediate states are absent.

To understand this behavior, we employ a micromagnetic model that solves the Landau-Lifshitz-Gilbert equation,

$$\frac{d\mathbf{m}}{dt} = -\gamma\, \mathbf{m} \times \mathbf{B}_{\text{eff}} + \alpha\, \mathbf{m} \times \frac{d\mathbf{m}}{dt} + \gamma B_{\text{SL}} \mathbf{m} \times \left(\mathbf{m} \times (\hat{\mathbf{z}} \times \hat{\mathbf{J}})\right) - \gamma B_{\text{FL}} \mathbf{m} \times (\hat{\mathbf{z}} \times \hat{\mathbf{J}}) \quad (1)$$

with $\gamma = 1.76 \times 10^{11} \text{s}^{-1}\text{T}^{-1}$ the electron gyromagnetic ratio, $\alpha$ the Gilbert damping constant, $\hat{\mathbf{J}}$ a unit vector along the current, and $\mathbf{B}_{\text{eff}}$ the effective magnetic field, which includes the external, exchange, anisotropy, demagnetizing, DMI and thermal[26] fields. The last two terms in (1) represent respectively the Slonczewski-like and field like SOTs[3,18,24], with amplitudes $B_{\text{SL}}$ and $B_{\text{FL}}$ proportional to the current density. The DMI effective field is given by[18,19,27]

$$\mathbf{B}_{\text{DMI}} = -\frac{2D}{M_S}\left[(\nabla \cdot \mathbf{m})\hat{\mathbf{z}} - \nabla m_z\right] \quad (2)$$

with $D$ the DMI constant. Note this expression only holds true at the interface, so it is applicable to thin films.

Simulation parameters correspond to the experimental sample dimensions $15000 \times 1200 \times 0.6$ nm, Gilbert damping constant $\alpha = 0.3$, saturation magnetization $M_S = 8.3 \times 10^5$ A/m, uniaxial out of plane anisotropy constant $K_U = 4.8 \times 10^5$ J/m$^3$ and exchange constant $A = 1.6 \times 10^{-11}$ J/m. The SHE angle and DMI constants are respectively taken from Refs. [5] and [25] as $\theta_{SH} = 0.07$ and $D = -1.2 \times 10^{-3}$ J/m$^2$ for Pt/CoFe/MgO, and $\theta_{SH} = -0.25$ and $D = -0.05 \times 10^{-3}$ J/m$^2$ for Ta/CoFe/MgO. $B_{SL}$ is computed as arising purely from the SHE via $B_{SL} = \frac{\hbar \theta_{SH}}{2|e|M_S t} J$, with e the electron charge, $\hbar$ Planck's constant, and $t$ the sample thickness. Unless specified otherwise, all simulations are performed at $T = 300$ K. Simulations were performed with GPMagnet[26], a GPU-based parallelized commercial code, using a sixth-order Runge-Kutta method with $5 \times 5 \times 0.6$ nm cell discretization and a 1ps computational time step.

Our first simulations aim to give insight into the influence of thermal fluctuations and DMI on switching dynamics. Here, we apply a continuously swept current at a ramp rate of $4 \times 10^{10}$ (A/m$^2$)/ns. With $D = T = 0$, we observe clean switching in which the magnetization remains uniform except for the effect of the demagnetizing field near the edges. In this case, a macrospin model suffices to reproduce the hysteresis loop precisely. As expected, increasing $T$ yields a narrowing of the hysteresis loop (Fig. 3a). In this case switching takes place via random nucleation of bubbles which expand throughout the whole sample (Fig. 3c), as described previously[24]. This non-uniform switching has however very little impact on the averaged $m_z$, yielding no remarkable effects aside from a decrease in the switching current.

When DMI is introduced, the magnetic switching becomes more irregular (i.e. the bubbles are distorted and expand non-isotropically) but the switching current and the averaged $m_z$ are barely affected for weak DMI (Fig. 3b). However, above a threshold DMI constant, which we find to be between 1 and 1.5 times $\sqrt{AK_U}$ (in agreement with ref. 28) the DMI is strong enough to promote non-uniform magnetization patterns. In particular, starting from a uniform state at zero current, and increasing the current, we observe the generation of a stripe-like structure, which breaks into a deformed skyrmion lattice, eventually disappearing for high enough currents; this effect is the responsible for the 'tails' in the hysteresis loops in Fig. 3b. Lastly, we find that the switching current decreases with decreasing current ramp rate.

We next seek to reproduce micromagnetically the pulsed switching behavior in Figs. 1c,d, considering 50 ns long pulses with 1 ns rise and fall times. In the Pt/CoFe sample simulations (Fig. 4a), we found that the switching is indeed not deterministic, and it starts via random nucleation of deformed skyrmions or random magnetization reversal on the edges of the sample. If the nucleation occurs within the intermediate switching region, the current intensity is insufficient to make this state vanish, and the non-uniformities spread throughout the sample. This is the cause of the intermediate states observed in Fig. 4a (red dashed circles); for larger currents the probability of non-uniform state generation is higher and the magnetization eventually switches during the pulse application, but it does not reach a fully uniform state. After the pulse is turned off, the non-uniformities spread, giving rise to the stabilization of these intermediate states. In this case, we observe a metastable helical magnetization configuration (Fig. 4c) in the form of stripe-like domains separated by left-handed Néel type DWs, as expected for strong DMI. The periodicity of the helix is ~200 nm, in agreement with the theoretical value[27] $4\pi A/D = 168$ nm. If nucleation occurs outside the intermediate switching region (for pulse amplitudes higher than the switching current density) we observe an isolated intermediate state (Fig. 4a, black circle), which reverts back to the uniform state after the application of a subsequent current pulse of sufficiently large intensity and length.

For the Ta/CoFe samples, experiments reveal a switching current density that is lower than for Pt/CoFe by a factor of ~10. This cannot be explained alone by the higher spin-Hall angle in Ta, nor can be attributed to the DMI, which is much weaker in Ta/CoFe/MgO than in Pt/CoFe/MgO[25]. Moreover, the anisotropy field and coercivity are nearly the same for these two samples. The explanation of this lower switching current threshold is instead the presence of a strong current induced field-like SOT, measured in Ref. 5 to be $B_{FL} \simeq 0.4$ T per $10^{12} A/m^2$, which is a factor of ~20 times larger than was measured for Pt/CoFe. In order to quantitatively reproduce the observed switching current values, we consider a larger $B_{FL} \simeq 2.5$ T per $10^{12} A/m^2$. Including this large field-like torque in the simulations leads the magnetization into the direction perpendicular to the injected current during the pulse, but after the pulse the magnetization precesses back to the uniform state, either 'up' or 'down' depending on the sign of the current.

The simulations for the Ta/CoFe sample (Fig. 4b) agree well with the experimental results. A remarkable aspect of the results in Fig. 4b is the absence of stable intermediate states in the magnetization[29] due to the much lower DMI[25]. In this case, the macrospin model perfectly reproduces micromagnetic results. Equally remarkable is the influence of the strong field-like SOT on the dynamical switching process; in its presence the nonuniform bubble domain nucleation observed in Fig. 3(c) is suppressed. In contrast to the case of the Slonczewski-like torque alone, here the magnetization switches in a quasi-uniform manner during the pulse without the formation of transient bubble states.

It is clear from this analysis that the switching process has a crucial stochastic component. The thermal agitation favors the nucleation of bubbles, which is greatly intensified when non-uniform magnetization is promoted by the DMI, and which can be suppressed by a strong field-like SOT. We can naively describe our sample as a tristable system; there are three possible states, 'up', 'down' and 'intermediate'. The DMI and exchange strength determine whether the 'intermediate' state is actually a metastable state or not, as well as the height of the energy barrier between this state and the uniform ones. The applied current modifies the energy barriers between the states via the Slonczewski-like torque, and the thermal agitation is responsible for jumps over these barriers (i. e. early switching processes).

In summary, we have experimentally studied SOT switching of heavy-metal/ferromagnet/oxide trilayers and for the first time evaluated micromagnetically the influence of the DMI and field-like SOT. We analyzed current-pulse-induced switching and observed the generation of stripes in the sample, an effect which is heavily dependent on temperature, the DMI strength and the switching time scale. Both experiments and simulations reveal the presence of helical magnetization intermediate states in current-pulse-induced switching in Pt/CoFe/MgO, while hysteresis loops in Ta/CoFe/MgO are sharp, demonstrating the crucial role of intermediate states stabilized by the DMI. Macrospin simulations satisfactorily reproduce the switching characteristics in the absence of DMI, but a micromagnetic approach is required to fully understand the processes that take place during the switching. Although the understanding of current-induced magnetization dynamics in these materials is still in its early stages, our results point out the possibility of stabilizing complex magnetization patterns such as helices or skyrmions which present promising perspectives for high-performance spintronics applications.

**Acknowledgements**

This work was supported by projects MAT2011-28532-C03-01 from the Spanish government and SA163A12 from Junta de Castilla y Leon. Work at MIT was supported by C-SPIN, one of the six SRC STARnet Centers, sponsored by MARCO and DARPA, and by the National Science Foundation under NSF-ECCS-1128439. Technical support from D. Bono is gratefully acknowledged.

[29]Applying a current pulse of magnitude close to the switching current density sometimes results in the appearance of two or more domains in the magnetization. However, unlike the intermediate states in Pt/CoFe/MgO, which are stable, these configurations always vanish to a uniform state after a time of the order of 1 μs.

**Figure captions**

**Fig. 1.** (a) Scanning electron micrograph of device. (b) Schematic of stack structure and experiment geometry. (c), (d) SOT switching for (c) Pt/CoFe, with $B_x = 50$ mT, showing non-deterministic switching and intermediate states, and (d) Ta/CoFe with $B_x = 10$ mT, showing sharp switching without intermediate states.

**Fig. 2.** Switching phase diagrams for (a) Pt/CoFe and (b) Ta/CoFe. (c), (d). Mean $m_z$ after current pulse application versus (c) current density at fixed $B_x$, and (d) versus $B_x$ at fixed current pulse amplitude for Pt/CoFe. Histogram in (e) shows distribution in $m_z$ for 50 switching cycles for fixed $B_x$ and $J$ on the switching boundary. (f) Mean $m_z$ after current pulse application versus $B_x$ at fixed current pulse amplitude for Ta/CoFe.

**Fig. 3.** (a) Influence of the temperature on switching with no DMI for a current ramp of $4 \times 10^{10}$ (A/m$^2$)/ns. (b) Influence of the DMI constant $D$ in the switching at $T = 300$K. Zero temperature macrospin results are also plotted. (c) Magnetization configuration during the rising current regime for $J = 0.44 \times 10^{12}$ A/m$^2$ with $T = 300$K and $D = 0$.

**Fig. 4.** (a) Simulated Pt/CoFe hysteresis loop for $B_x = 50$ mT using a macrospin model (dotted line) and computed micromagnetically, showing the presence of intermediate states near the switching transition (red dashed circles) and an isolated intermediate state outside the switching transition (black circle). (b) Corresponding simulations for Ta/CoFe, showing sharp switching transition and significantly lower switching current. (c) Exemplary helical magnetization intermediate state in Pt/CoFe. Red means up magnetization and blue means down magnetization. This configuration is stable in the absence of current over arbitrarily long times.

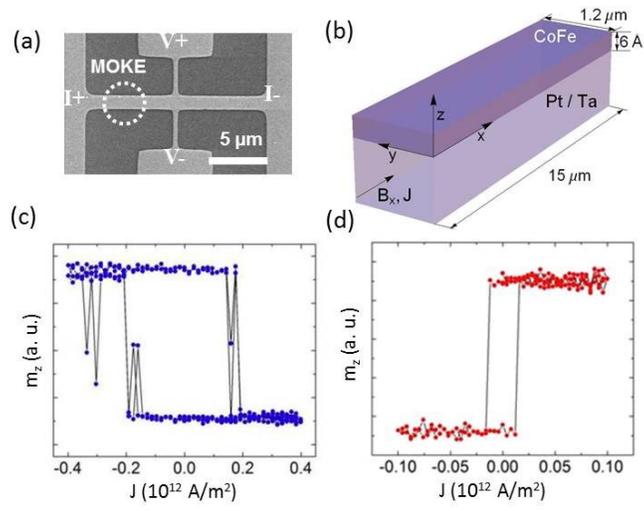

**Fig. 1.**

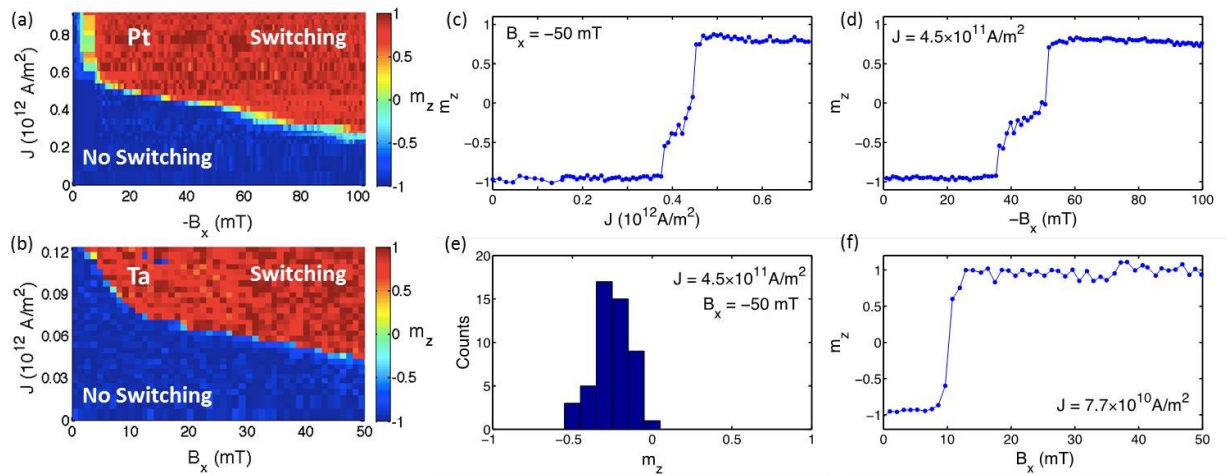

**Fig. 2.**

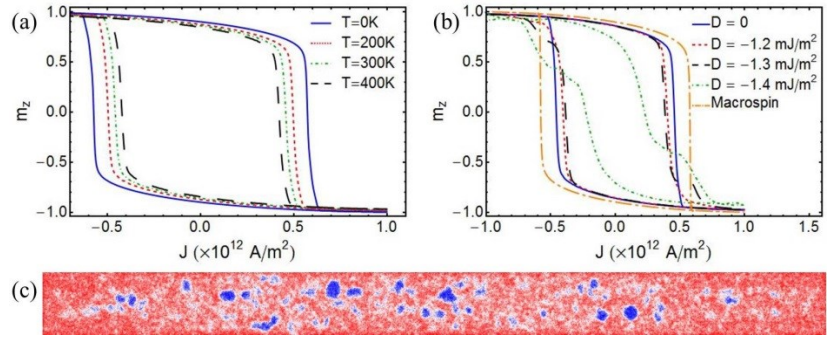

**Fig. 3.**

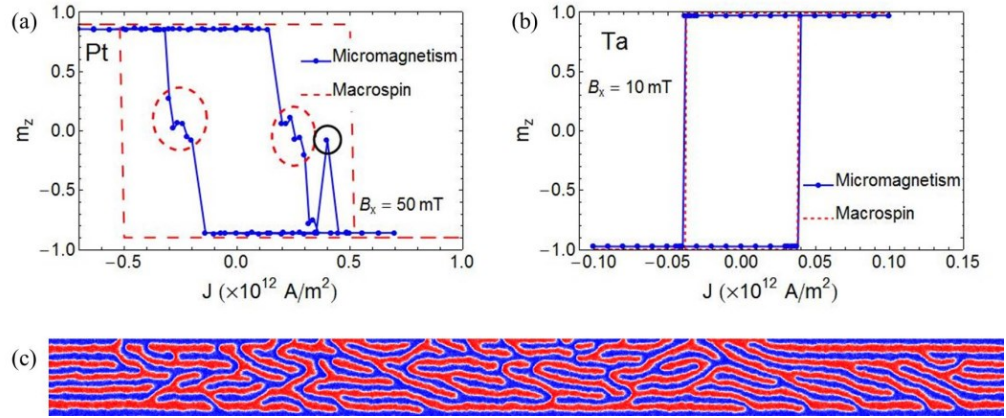

**Fig. 4.**